\documentclass[prl,twocolumn,amsmath,amssymb]{revtex4}
\usepackage{graphicx}
\usepackage{color}
\usepackage{graphicx}% Include figure files
\usepackage{dcolumn}% Align table columns on decimal point
\usepackage{bm}% bold math
\usepackage{amsmath}
\usepackage{amsfonts}
\usepackage{bbm}
\usepackage{subfigure}
\usepackage{setspace}

\usepackage{pstricks}
\usepackage{color}

\newcommand{\beq}{\begin{eqnarray}}
\newcommand{\eeq}{\end{eqnarray}}

\newcommand{\bmp}{\noindent\begin{minipage}{16cm}}
\newcommand{\emp}{\end{minipage}\vskip 7mm} % 7mm untightened

\def\drawbox#1#2{\hrule height#2pt
        \hbox{\vrule width#2pt height#1pt \kern#1pt
              \vrule width#2pt}
              \hrule height#2pt}

\def\Asym#1#2{\vcenter{\vbox{\drawbox{#1}{#2}
              \kern-#2pt % line up boxes
              \drawbox{#1}{#2}}}}

\def\mpt{{\slash\!\!\!\!\!\:p}_T}

\begin{document}
%%%%%%%%%%%%%%%%%%%%%%%%%%%%%%%%%%%%%%%%%%%%%%%%%%%%%%%%%%%%%%%%%%%%%%%%%%%
\title{\Large  \color{red}  Technicolor Dark Matter}
\author{Roshan {\sc Foadi}}
\email{foadi@ifk.sdu.dk}
\author{Mads T.{\sc Frandsen}}
\email{toudal@nbi.dk}
\author{Francesco {\sc Sannino}}
\email{sannino@ifk.sdu.dk}
\affiliation{Center for High Energy Physics, University of Southern Denmark, Campusvej 55, DK-5230 Odense M, Denmark.}

%%%%%%%%%%%%%%%%%%%%%%%%%%%%%%%%%%%%%%%%%%%%%%%%%%%%%%%%%%%%%%%%%%%%%%%%%%%%%%%%%%%%%%%%%%%%%%%%%%%%%%%%%%%%%%%%%%%%%%%%%%%%%%%%%%%%%%%%%%%%%%

%%%%%%%%%%%%%%%%%%%%%%%%%%%%%%%%%%%%%%%%%%%%%%%%%%%%%%%%%%%%%%%%%%%%%%%%%%%%%%%%%%%%%%%%%%%%%%%%%%%%%%%%%%%%%%%%%%%%%%%%%%%%%%%%%%%%%%%%%%%%%%

\begin{abstract}
Dark Matter candidates are natural in Technicolor theories. We introduce a general framework allowing to predict signals of Technicolor Dark Matter at colliders and set constraints from earth based experiments such as CDMS and XENON. We show that the associate production of the composite Higgs can lead to relevant signals at the Large Hadron Collider.\end{abstract}

%%%%%%%%%%%%%%%%%%%%%%%%%%%%%%%%%%%%%%%%%%%%%%%%%%%%%%%%%%%%%%%%%%%%%%%%
\maketitle

Decaying Dark Matter (DM) models \cite{Nardi:2008ix} find a natural setting within recent Technicolor (TC) extensions of the Standard Model (SM) summarized in \cite{Sannino:2008ha}. The way to the new phenomenologically viable TC models was opened in \cite{Sannino:2004qp} and the conjectured beta function for nonsupersymmetric gauge theories  \cite{Ryttov:2007cx}  is a useful tool to  further enlarge the number of TC models \cite{Ryttov:2008xe}.   
In these models the lightest technibaryon can be identified with the DM candidate \cite{Nussinov:1985xr,Barr:1990ca,Gudnason:2006yj}. In \cite{Nardi:2008ix} it was shown that decaying DM models can fit the excess of Cosmic Ray $e^{\pm}$ spectra measured by PAMELA \cite{Adriani:2008zr} and ATIC \cite{ATIC-2}.  These experimental results are in agreement with other  recent observations \cite{Adriani:2008zq,Collaboration:2008aa}. What is encouraging is that decaying DM  models able to explain the  PAMELA and ATIC excess are also in natural agreement with the constraints from the spectrum of gamma rays \cite{Nardi:2008ix}  studied by the HESS collaboration \cite{:2003gg,Aharonian:2007km}. The latter poses a challenge for single component DM models where the relic density is due to an annihilation cross section \cite{Bertone:2008xr,Regis:2008ij,Bell:2008vx}. In \cite{Nardi:2008ix} it was observed that the decay rate for the DM is consistent with a Grand Unified picture within TC models  \cite{Gudnason:2006mk}. A similar observation has been made in  \cite{Arvanitaki:2008hq} for supersymmetric extensions of the SM.

Here we provide the first investigation of the collider phenomenology coming from the dark sector  of new strongly coupled dynamics. Since the PAMELA and ATIC observations could still have an astrophysical origin  we consider a more general parameter space especially relevant for LHC phenomenology.

We indicate with TIMP a generic Technicolor Interacting Massive Particle which is stable at low energies.  Examples are the lightest technibaryon as well as any other composite state protected against decay  by a symmetry. These states occur, for example,  in Ultra Minimal Walking Technicolor  (UMT) \cite{Ryttov:2008xe}. Other DM candidates related to TC have been investigated in \cite{Kainulainen:2006wq,Kouvaris:2007iq}. Assuming that one of the TIMPs is the DM particle, then constraints from CDMS \cite{Ahmed:2008eu} and XENON \cite{Angle:2007uj} -- under the assumption of a single component DM -- strongly favor TIMPs  which are overall neutral under the electroweak interactions. 

{}We identify the TIMP with a complex scalar $\phi$, singlet under the SM interactions, charged under the $U(1)$ technibaryon symmetry of a generic TC theory. For example UMT includes such a state \cite{Ryttov:2008xe}.  The main differences with other models featuring an extra $U(1)$ scalar DM are: i)  The $U(1)$ is natural, i.e. it is identified with a technibaryon symmetry; ii) Compositeness requires the presence around the electroweak scale of spin one resonances leading to striking collider signatures; iii) The DM relic density is naturally linked to the baryon one via an asymmetry. 
We also introduce a (light) composite Higgs particle since it appears in several  strongly coupled theories as shown in the Appendix F of \cite{Sannino:2008ha}. 
 We consider only the TC global symmetries relevant for the electroweak sector, i.e. the  $SU(2)\times SU(2)$ spontaneously breaking to $SU(2)$. The low energy spectrum contains, besides the composite Higgs, two $SU(2)$ triplets of (axial-) vector spin one mesons.  The effective Lagrangian, before including the TIMP, has been introduced in \cite{Foadi:2007ue,Belyaev:2008yj} in the context of minimal walking theories. Here we add the TIMP  $\phi$ with the following interaction terms:
\begin{widetext}
\begin{eqnarray}
{\cal L}_{\rm DM}&=&\frac{1}{2}\ \partial_\mu\phi^\ast\partial^\mu\phi-\frac{M_\phi^2 - d_M\ v^2}{2}\ \phi^\ast\phi
+\frac{d_F}{2\Lambda^2}\ {\rm Tr}\left[F_{{\rm L}\mu\nu} F_{\rm L}^{\mu\nu}+F_{{\rm R}\mu\nu} F_{\rm R}^{\mu\nu}\right]\ \phi^\ast\phi  \nonumber \\
&+& 
d_C\ \tilde{g}^2 \ {\rm Tr}\left[C_{{\rm L}\mu}^2+C_{{\rm R}\mu}^2\right]\ \phi^\ast\phi
-\frac{d_M}{2}\ {\rm Tr}\left[M M^\dagger\right]\ \phi^\ast\phi \ .
\label{eq:Lagrangian}
\end{eqnarray}
\end{widetext}
The matrix $M$ contains the composite Higgs and the pions eaten by the electroweak bosons,
\begin{eqnarray}
M &=& \frac{1}{\sqrt{2}}\left[v+H+2\ i\ T^a\ \pi^a\right] \ ,
\end{eqnarray}
where $v$ is the vacuum expectation value, and $T^a=2\sigma^a$, $a=1,2,3$, where $\sigma^a$ are the Pauli matrices. $\Lambda $ is a scale associated to the breakdown of the low energy effective theory and it is in the TeV energy range.  $C_{{\rm L}\mu}$ and $C_{{\rm R}\mu}$ are the electroweak covariant linear combinations $\displaystyle{
C_{{\rm L}\mu} = A^a_{{\rm L}\mu}\ T^a-{g}/{\tilde{g}}\ \widetilde{W}^a\ T^a }$ and $\displaystyle{
C_{{\rm R}\mu} = A^a_{{\rm R}\mu}\ T^a-{g^\prime}/{\tilde{g}}\ \widetilde{B}\ T^3}$, 
where $\widetilde{W}^a$ and $\widetilde{B}$ are the electroweak gauge fields (before diagonalization).  $F_{\rm L}$, $F_{\rm R}$ are the fields strength tensors associated to the vector meson fields $A^a_{{\rm L}\mu}$ and $A^a_{{\rm R}\mu}$, respectively. The associated spin one mass eigenstates are indicated with $R_{1}$ and $R_2$.
  
  \vskip .3cm
\noindent
{\bf Earth Based Constraints} 
\vskip .2cm
In earth based experiments the TIMP interacts  with nuclei mainly via the exchange of a composite Higgs (the $d_M$ term in the above Lagrangian) and a photon. The photon interaction was considered in \cite{Bagnasco:1993st} and it is due to a nonzero electromagnetic charge radius of $\phi$. Here we stress the relevance of the composite Higgs exchange. The Lagrangian term for the charge radius interaction is:
\beq 
\mathcal{L}_B=i e \frac{d_B}{\Lambda^2} \phi^*\overleftrightarrow{\partial_\mu} \phi \, \partial_{\nu}F^{\mu\nu} \ .
\label{eq:Bagnasco}
\eeq
 The result for the non-relativistic cross section per nucleon for scattering off a nucleus is:
\begin{equation}
\sigma_{\textrm{nucleon}} = \frac{\mu^2}{4\pi} \left[ \frac{Z}{A} \frac{ 8\pi \, \alpha \, d_B}{\Lambda^2} + 
\frac{d_M  \, f m_N}{M^2_H M_{\phi}}\right]^2 \ ,
\end{equation}
with $A$ the mass number, $Z$ the atomic one and $\mu= M_{\phi} m_N /(M_{\phi} + m_N)$.
Here $m_N$ is the mass of the nucleon, and $f\sim 0.3$ parametrizes the Higgs to nucleon coupling \cite{Shifman:1978zn}.  The composite Higgs can be light in theories with higher dimensional representations \cite{Sannino:2008ha,Hong:2004td} as well as  near-conformal TC models \cite{Dietrich:2005jn}. Similarly $\phi$ can also be light~\cite{Ryttov:2008xe}. These observations justify keeping both contributions for the cross section. 

\begin{figure}[htp!]
{\includegraphics[height=6cm,width=8.5cm]{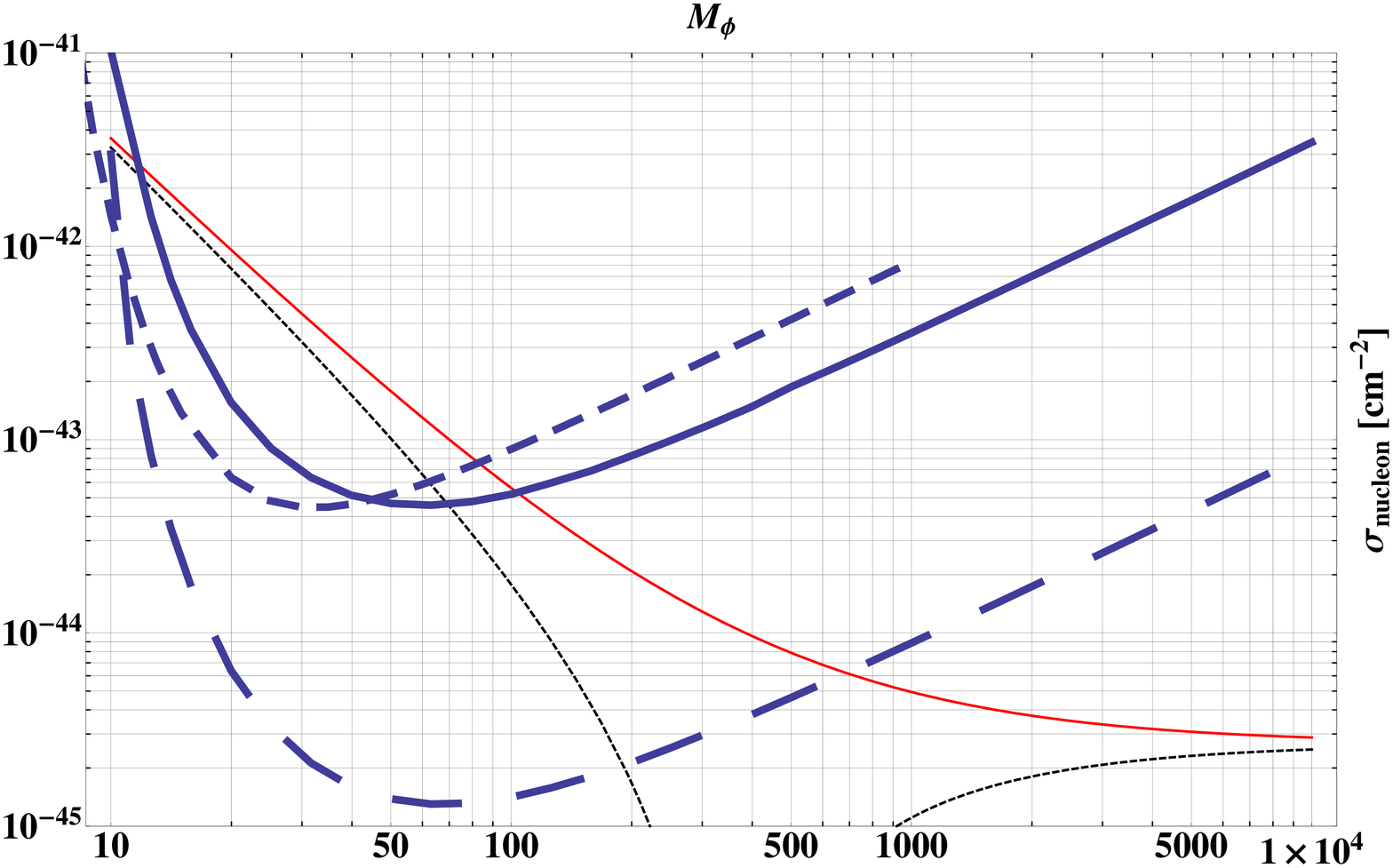}}
\caption{ TIMP -- nucleon cross section  for $d_M=1$, $\Lambda=3$ TeV, $M_H=300$ GeV and $d_B=\pm 1$ as a function of the TIMP mass. The plus sign corresponds to the thin-solid-red line and the minus one to the thin-dotted-black line. Also plotted is the approximate exclusion limit from CDMS II Ge Combined  (solid-thick-blue) and XENON10 2007 (dashed-thick-blue).  The long-dashed-blue line is the projected superCDMS exclusion curve. The allowed region is below the CDMS and XENON curves.}\label{fig:cdms}
\end{figure}

In figure~\ref{fig:cdms} we plot the cross sections for a reference value of $M_H = 300$ GeV, $\Lambda = 4\pi  F_{TC} \approx 3 $ TeV, $d_M=1$ and $d_B=\pm 1$. In the same plot we compare our curves with the CDMS (solid-thick-blue) and XENON (dashed-thick-blue) exclusion limits. The long-dashed-blue line is the projected superCDMS exclusion curve \cite{Jeff}. 

At large $M_\phi$  the single photon exchange dominates and in this region, for the parameters chosen, the associated cross section is well below the experimental limits. At low values of the TIMP mass the Higgs exchange dominates and the experimental constraints become relevant.

\begin{figure}[htp!]
{\includegraphics[height=4cm,width=6cm]{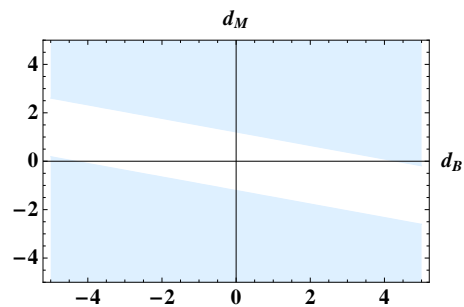}}
\caption{CDMS exclusion limit in the $d_B - d_M$ plane for a 100 GeV TIMP with $\Lambda=3$ TeV and $M_H=300$ GeV. The colored region is excluded. }\label{fig:exclusion}
\end{figure}

The deep around $500$ GeV for the dashed-black line is due to destructive interference between the Higgs and the single photon exchange terms. 
It is useful also to plot, in Fig.~\ref{fig:exclusion}, the CDMS exclusion limits in the $d_B - d_M$ plane. We choose a $100$ GeV TIMP and a mass of the composite Higgs of $300$ GeV. The allowed region is the uncolored one.

\vskip .3cm
\noindent
{\bf Collider Signals of Technicolor Dark Matter} 
\vskip .2cm
If the TIMP mass is of the order of a TeV, as expected in a traditional TC model, it will not be produced at the LHC. Interesting phenomenology at the LHC will occur, however, if the  TIMP has a mass of the order of hundreds of GeVs. Models featuring such a light TIMP are, for example: i) Multi scale TC \cite{Lane:1989ej} in which the TC scale is lowered with respect to the electroweak one; ii) Models in which the technibaryon is a pseudo Goldstone boson \cite{Ryttov:2008xe}. 

{}From the LHC point of view it is very interesting to explore the possibility of a TIMP lighter than the composite Higgs.  A TIMP lighter than half of the Higgs mass will result in invisible decays of the Higgs itself \cite{Shrock:1982kd}.  A new ingredient with respect to earlier studies is constituted by  the interplay between the TIMP and the spin one resonances coming from the strong dynamics per se. 

\begin{figure}[htp!]
{\includegraphics[height=3cm,width=7.5cm]{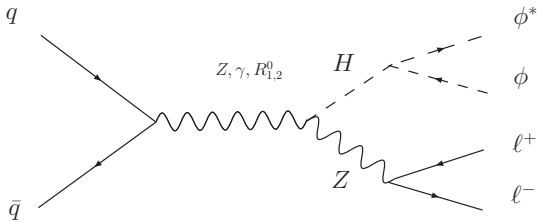}}
\caption{Feynman diagram for missing energy signal in a TC model with (axial-) vector resonances $R_{1,2}$ and a TIMP $\phi$.}\label{fig:dmsignal}
\end{figure}

We focus on the LHC process \cite{Godbole:2003it,Davoudiasl:2004aj} shown in Fig.~\ref{fig:dmsignal}:
\beq 
pp\to Z H \to \ell^+ \ell^- \mpt \ ,
\eeq
where the missing transverse momentum ($\mpt$) signal arises from the invisible decay of the Higgs into $\phi^\ast \phi$.
In \cite{Belyaev:2008yj} we have shown that the light composite Higgs production in association with a SM gauge boson is enhanced when a light axial spin one state is also present in the low energy spectrum.  Walking Technicolor models can accommodate such a spectrum \cite{Appelquist:1998xf,Dietrich:2005jn,Foadi:2007ue}.

 The acceptance cuts relevant for LHC are
\begin{eqnarray}
|\eta^{\ell}|<2.5\ , \quad  p_T^\ell> 10 \mbox{ GeV} \ , \quad   \Delta R(\ell\ell) > 0.4  \ .
\label{eq:cuts1}
\end{eqnarray}
Here $\ell$ is a charged lepton, $\eta^{\ell}$ and $p_T^\ell$ are the pseudo-rapidity and transverse momentum of a single lepton while $\Delta R$ measures the separation between two leptons in the detector. $\Delta R$ is defined via the difference in azimuthal angle $\Delta\phi$ and rapidity $\Delta\eta$ between two leptons as $\Delta R\equiv \sqrt{(\Delta\eta)^2+(\Delta\phi)^2}$. 

The main sources of background come from di-boson production followed by leptonic decays
\beq
ZZ \to \ell^+\ell^- \nu \bar{\nu} \ , \ W^+ W^- \to \ell^+ \nu \ell^-\bar{\nu} \ , \ ZW \to \ell^+\ell^- \ell \nu 
\eeq
where in the last process the lepton from the W decay is missed.
 
 We impose the additional cuts
\begin{eqnarray}
 |M_{\ell \ell}-M_Z|<10 \mbox{ GeV} \ , \quad {\rm and} \quad \Delta \phi(\ell\ell) < 2.5  \ .
\label{eq:cuts2}
\end{eqnarray}
The first is meant to reduce the WW background by requiring the invariant mass of the lepton pair to reproduce the Z boson mass. The second cut on the azimuthal angle separation together with taking large $\mpt$  reduces potential backgrounds such as single Z production + jets with fake $\mpt$ \cite{Godbole:2003it,Davoudiasl:2004aj}.

We generate the signal and  ZZ as well as WW backgrounds using the LanHEP/CalcHEP~\cite{Pukhov:2004ca,Semenov:2008jy} implementation of Lagrangian  \eqref{eq:Lagrangian}  which is an extension of \cite{Belyaev:2008yj}. 

\begin{figure}
{\includegraphics[height=5cm,width=7.5cm]{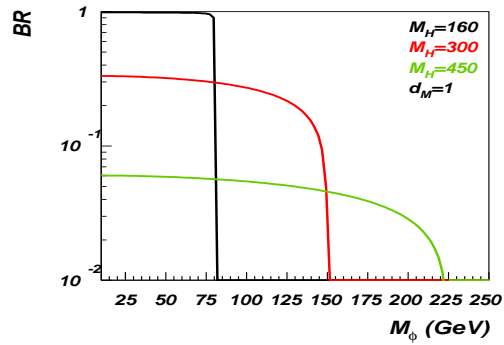}}
\caption{The invisible branching ratio of the Higgs into $\phi^\ast \phi$ for $d_M=1$ and $M_H=$160 (black), $M_H=$300 (red) and $M_H=450$ (green).}\label{fig:br}
\end{figure}

The invisible branching ratios of the composite Higgs are summarized in Fig.~\ref{fig:br}. They depend on the coupling $d_M$ and the TIMP mass $M_{\phi}$. 

\begin{figure*}[htp!]
{\includegraphics[height=7cm,width=7cm]{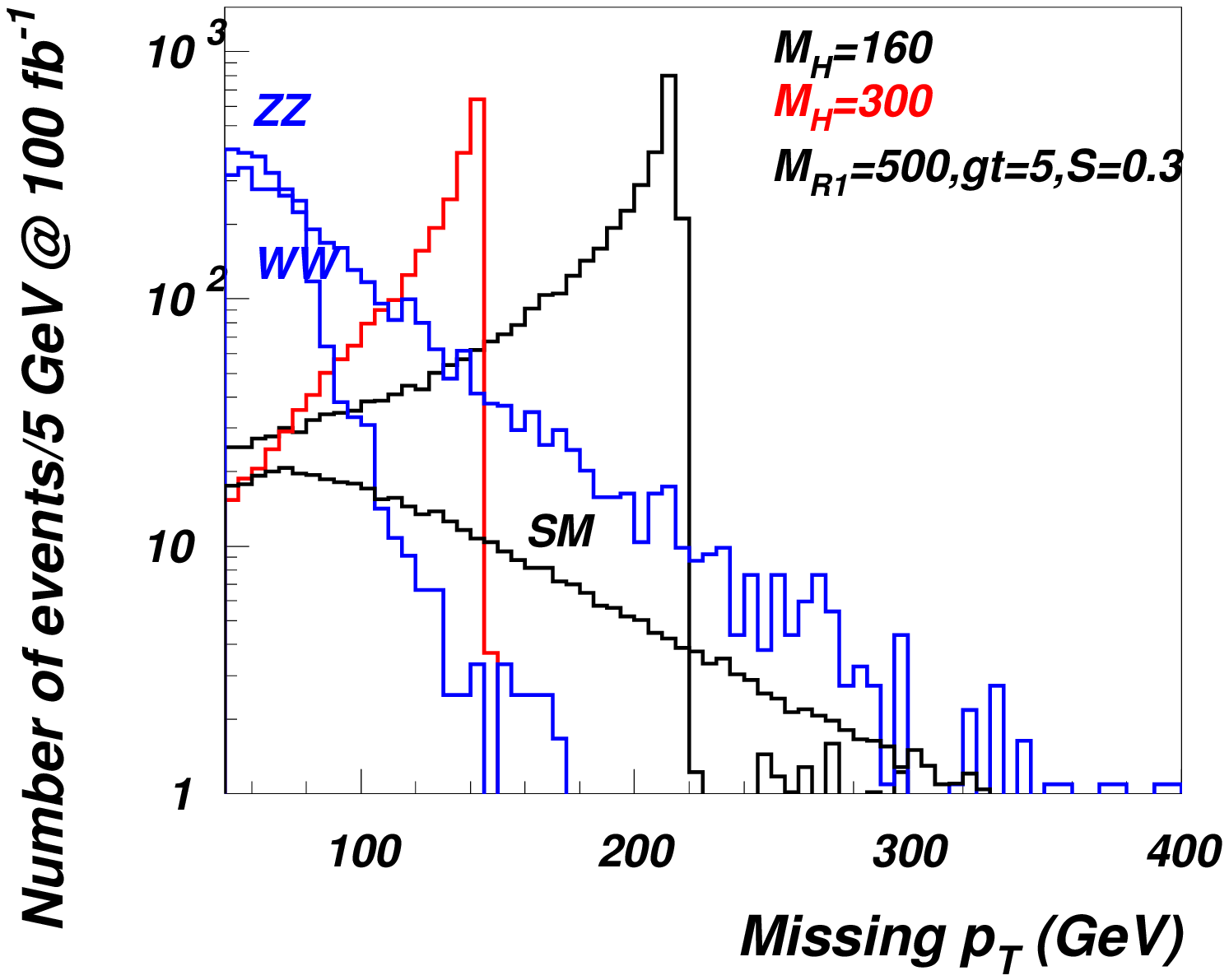}\hskip 1.5cm \includegraphics[height=7cm,width=7cm]{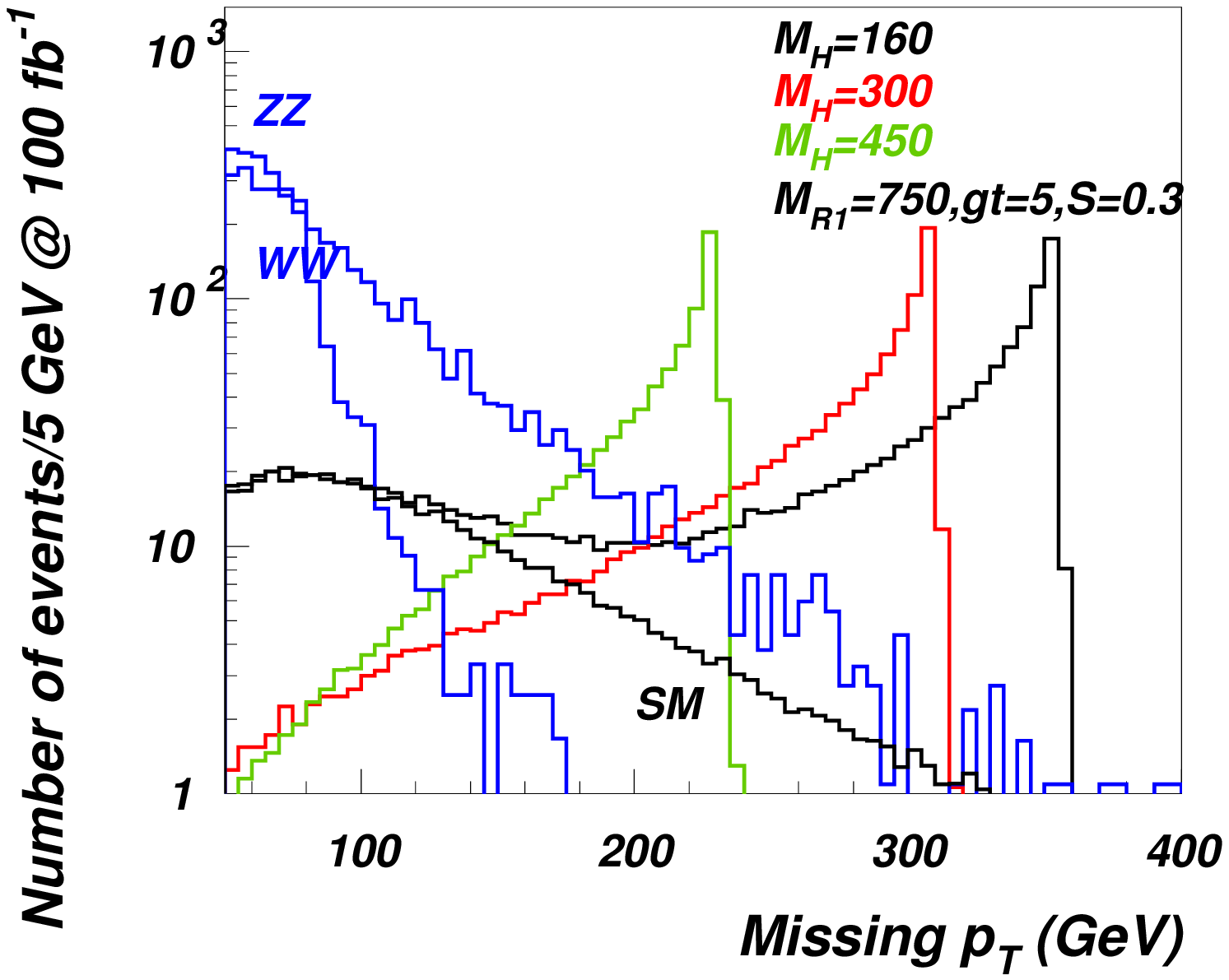}}
\caption{$pp \to ZH$ with $Z\to e^+ e^-$ for three different values of the composite Higgs mass. On the x-axis we plot  $\mpt$  assuming a completely invisible Higgs decay. Also shown (in the lower black curve) is the distribution arising from a SM Higgs with $M_H=160$. The cuts are those in Eq.~\ref{eq:cuts1} and~\ref{eq:cuts2}.}\label{fig:a}
\end{figure*}

Finally in figure~\ref{fig:a} we plot our predictions for $
pp\to Z H \to \ell^+ \ell^- \mpt $. The peaks correspond to different values of the composite Higgs mass, i.e. $160$, $300$, and $450$~GeV. In the left panel the peak associated with the $450$~GeV Higgs is well inside the background and located at lower values of the transverse momentum.  For comparison we plot the SM distribution for a $160$~GeV Higgs Mass. We also show the WW and ZZ backgrounds for this signal, which are identical for SM and TC. The backgrounds and the SM signal are in agreement with the distributions presented in \cite{Davoudiasl:2004aj} based on  MadGraph. The two plots in Fig.~\ref{fig:a} correspond to two distinct values of the $R_1$ (mostly axial state) mass .

The TC signals in the plots assume a completely invisible composite Higgs decay. To obtain the LHC signal one has to multiply each TC curve by the corresponding branching ratio plotted in Fig.~\ref{fig:br}.
 
To make the plots in Fig.~\ref{fig:a} we have chosen two reference values of the axial vector  mass $R_1$, $M_{R_1} \sim 500$  and $750$~GeV as well as $\tilde{g}=5$. {}For a light axial resonance such a relatively large value of $\tilde{g}$ is favored by unitarity arguments \cite{Foadi:2008xj,Foadi:2008ci}. We have shown in \cite{Foadi:2007ue} that it is possible to use the $S$ parameter to get information on the vector versus axial spectrum in walking TC theories. This allows us to trade the vector mass for the value of  Peskin-Takeuchi S parameter \cite{Peskin:1990zt} quoted in the plots. 

The results show that the combination of a light axial and composite Higgs yields a missing transverse momentum distribution from the $ZH$ production which is significantly different from that in the SM, giving rise to a potentially neat signal of a TIMP at colliders.

Summarizing, we investigated the constraints from earth based experiments on physically relevant TC type DM and for the first time investigated its potential for discovery at the LHC.

\end{document}